\documentclass[12pt]{article}
\usepackage[dvips]{graphicx}
\begin{document}

\title{Monodisperse approximation as a tool
to determine
stochastic effects in decay of
metastable phase}
\author{Victor Kurasov}

\maketitle

\begin{abstract}
Stochastic features of
decay of a metastable
phase have been
investigated with the
help of a new
monodisperce
approximation. This
approximation is more
precise than the already
used one and namely it
allows to give a very
simple but rather
precise way to calculate dispersion
of the total droplets number
initiated by stochastic appearance
of supercritical embryos. the
derivation is done for a free
molecular regime of droplets growth
but the diffusion regime is also
discussed.
\end{abstract}

The pioneer
investigations by Wilson  \cite{Wilson} more
than hundred years
ago opened systematic investigations of
nucleation. But only in the last years the
progress in
theoretical description of nucleation
allowed to put the
questions of stochastic properties of
 the first order phase transition kinetics.

Since \cite{Koll}, \cite{VestGr},
\cite{Kolldyn}
considerations of stochastic
properties of kinetics of
the first order phase transition can
be regarded
as one of the actual
problems to investigate.

The main statistic characteristics of the nucleation
process
are the mean value of the total number of droplets
$N_{tot}$
and dispersion of $N_{tot}$.
The mean
value $\bar{N}_{tot} = < N_{tot} > $
averaged
over all fluctuations and consequences
of their influence on  kinetics
 doesn't generally differ from a value
given by a theory of the averaged characteristics
(TAC) \cite{PhysRevE},
\cite{statiae}. This property is
explained in \cite{explanation} and
partially in \cite{statiae}, the
derivations in \cite{Koll} are illegal
since they are based on an
inappropriate linearization
(explanation see in \cite{statiae}).
As for the value of dispersion of
$$
\Delta  =
< (N_{tot} - <N_{tot}>)^2 >
$$
 one can not say that it is
found absolutely correctly, the deviation
between numerical simulations and results
 of \cite{Koll} is
about 10 percent, the error in \cite{statiae} is
practically absent, but  when the procedure from
\cite{statiae}
is applied to other regimes of droplets
growth (not for the free
molecular one)  the error can increase
up to 10 percent. So, the task to propose a way to
calculate dispersion remains rather actual.

One has to choose the
style of determination of dispersion
of the nucleation process.
The traditional way is to start
with the iteration solution and
to determine the dispersion on
the base of iteration. But this way can not lead to a
suitable result. Really, the deviation of
the square of dispersion  $D$ from
the standard value
$$
D_{standard} = 2 N_{tot}
$$
can be attained only by reaction
of the formation of new
droplets at the
latest moments of nucleation period on the
excess of the
droplets formed  in the first moments of
nucleation period.
But the reaction of the formation of
new droplets on formation
of other already existing
droplets appears only
in the second iteration in frames
of standard iteration
procedure (see \cite{Novosib}).
The second
iteration can not
be analytically calculated even
in the theory with
averaged characteristics, only the
number of droplets in
 can be determined analytically.

So, one has to come to
some method of calculation of
dispersion which is
based on some model behavior of
supersaturation.
An approach of such type was used in
\cite{Koll}, \cite{VestGr}.
In the cited papers the
approximation
proposed in \cite{book1} was used to calculate
the stochastic effects.
This approximation is the following:
during the first half
of nucleation period the droplets
are formed under the
ideal supersaturation, later all
remaining droplets are
formed under the vapor consumption by
the droplets from the first half.
This approximation
originally was
used in \cite{book1} only
for some rough estimates necessary
for justification of strong
inequalities necessary to
construct the mathematical models of kinetics.

The mentioned
model belong to the class of models with a
fixed boundary.
Namely the boundary between the first half
and the second
half has a fixed value - the transition
between cycles occurs in the fixed moment of time.

One can see that the two cycle models
with a fixed boundary
aren't  too suitable
for calculation of stochastic
properties of the phase
transition kinetics. Really, the
stochastic deviations of
characteristics of the first cycle
means that in the prescribed
approximation the parameters
coming from the evolution
before the boundary
will differ from the
value in the theory based on the
averaged values.

But the model used in
investigation of stochastic effects
was chosen  namely for the theory with
averaged
characteristics. Why shall it work with other
values of parameters?
Certainly, there is no reason for
applicability of
the model in such situation. This is the
main contradiction
in the application of two cycle model
with a fixed boundary
in investigation of stochastic
properties.

Fortunately the situation of decay
in a free molecular  regime of droplets growth
can be roughly described on the
base of the model with a
fixed boundary. This possibility is explained by
existence of specific zone -
the buffer zone \cite{statiae}.
The description of stochastic
effects on base of the
mentioned approximation requires some
rather complex constructions
because the buffer cycle
complicates construction. The precision
of calculations isn't too high
because the buffer zone isn't too  long and can
be extracted
with a certain
imagination.

Alternative
possibility is to use the model with a floating
boundary. In these models
the boundary is determined from
equations corresponding to the attaining of some
values of some characteristics of process.
The model of monodisperce
spectrum
\cite{Monodec} belongs to this class.

The property of internal time in the kinetics of
decay observed in \cite{statiae},
\cite{varios} states that the system can
wait so long as possible for appearance
of essential quantity
of droplets in the system and
nothing will be changed in
nucleation kinetics.
The process will simply be shifted in
the time scale.

In the model of monodisperse spectrum
with  a floating boundary
the number of droplets appears as
the determining characteristic.
In our therminology the determining parameter is the
characteristic which fixed value has to
be attained at the
boundary.

What type of model
with the floating boundary we have to
choose?
Certainly, one can
propose the simple generalization  of
the model used in \cite{statiae}:
until the boundary the
rate  of
appearance of droplets
is ideal one and later the evolution
is governed by droplets
appeared before the boundary.
If we suppose that the
rate of droplets formation is ideal
(unperturbed by the
vapor consumption) until some moment,
then we have to
take into account that the
average amplitude of
spectrum  for the droplets
formed in the first cycle is
one and the same for all of them.
Then the subintegral
function in expression for $g$ has a special power
behavior.
Namely this behavior is the base to determine the
boundary (not only
the number of droplets in the first cycle
determines the
boundary).
At least  one has to take into account
all four first momenta of the spectrum.
So, on one
hand the theory becomes very complex.
On the other hand
the constant value of spectrum in the
first cycle is
violated by stochastic fluctuations and the
approximation can
be violated also.
This is a certain disadvantage of th emodel.
So, this model can
not be effectively used.

Also here appears a question:
what characteristic will
determine the boundary?
Or one can put a question: what
equation on the boundary
should be written? There is no
clear answer on these questions.

So, the simplest model
satisfying our requirements is the
monodisperse model.
Here we don't face such
difficulties as in
the models described previously
because namely the number of droplets stands
both in monodisperse approximation
and in Gaussian
distribution for the
number of droplets formed during the
first cycle.
Also the number of droplets is the determining
characteristic here.

Recall briefly the monodisperse
approach. The evolution
equation after
rescaling can be written as \cite{Monodec}
\begin{equation} \label{tt}
g(z) = \int_0^z (z-x)^3 \exp(-g(x)) dx
\end{equation}
where unknown
function $g$ is the renormalized  value of
the number of molecules in a new phase.

In monodisperce approximation
one can use for $g$ the model
where all "essential" droplets
have one and the same size,
i.e.
$$
 g(z) \sim z^3
 $$

The number of essential
droplets $N_{ess}$ is parameter of
the model.

The total number of
droplets $N_{tot}$ is calculated as
$$
N_{tot} \sim \int_0^{\infty}
\exp(-g(x)) dx
$$

Precise solution of (\ref{tt}) gives
$$
N_{tot} = 1.28
$$
In the standard monodisperse approximation
$$
g(z)  =  N_{ess} z^3
$$
$$
N_{ess}  = 1/4
$$
and
$$
N_{tot} =
\int_0^{\infty} \exp(-x^3 /4) dx \approx 1.25
$$

We shall start  with  numerical
results for the number of
droplets formed in the process of decay.
The results for the total number of droplets
are drawn in
Figure 1.


\begin{figure}[hgh]

\includegraphics[angle=270,totalheight=10cm]{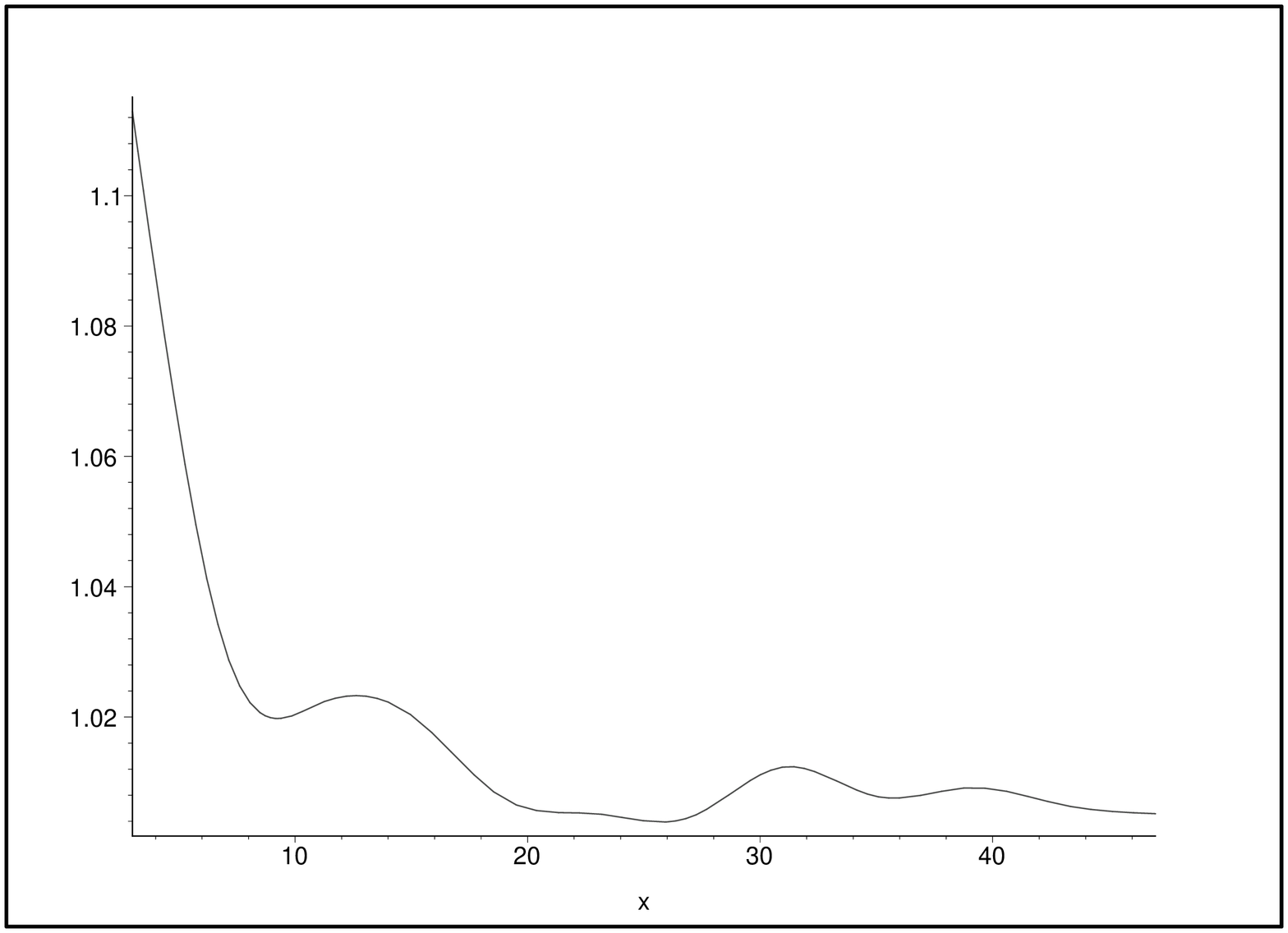}

\begin{caption}
{
Numerical solution.
The number of droplets in the units of
the number predicted by the mean values
theory as a function
of the volume of the system (or the
number of droplets in MVT).
}
\end{caption}
\end{figure}

One can note here that practically immediately
after the initial moment of time the number of
droplets is the mean value predicted by the
mean values
theory (MVT).
This isn't valid only for the first few droplets.

Now we shall present results in the
mean values of droplets
for the model with monodisperse
spectrum. Here the vapor
consumption occurs by the monodisperse
peak $N(\Delta x /4)$
as it prescribed by the monodisperse
variant of MVT
\cite{Monodec}.

One can prove that in all monodisperse
models the mean number of droplets
coincides with the corresponding values
calculated in MVT.

At this step there appears  a
little discrepancy because the monodisperse
spectrum in
the monodisperse variant of MVT begins to act at
the very beginning and we have to wait
until the moment
when $N_{tot}/4 = 1/4$ will appear.
So, a little shift in
the mean value of droplets predicted
by MVT appears. Since we draw
relative numbers,  i.e.
$<N> / N_{tot}(V=\infty)$ there
will be no error.

The results are drawn in Figure 2.


\begin{figure}[hgh]

\includegraphics[angle=270,totalheight=10cm]{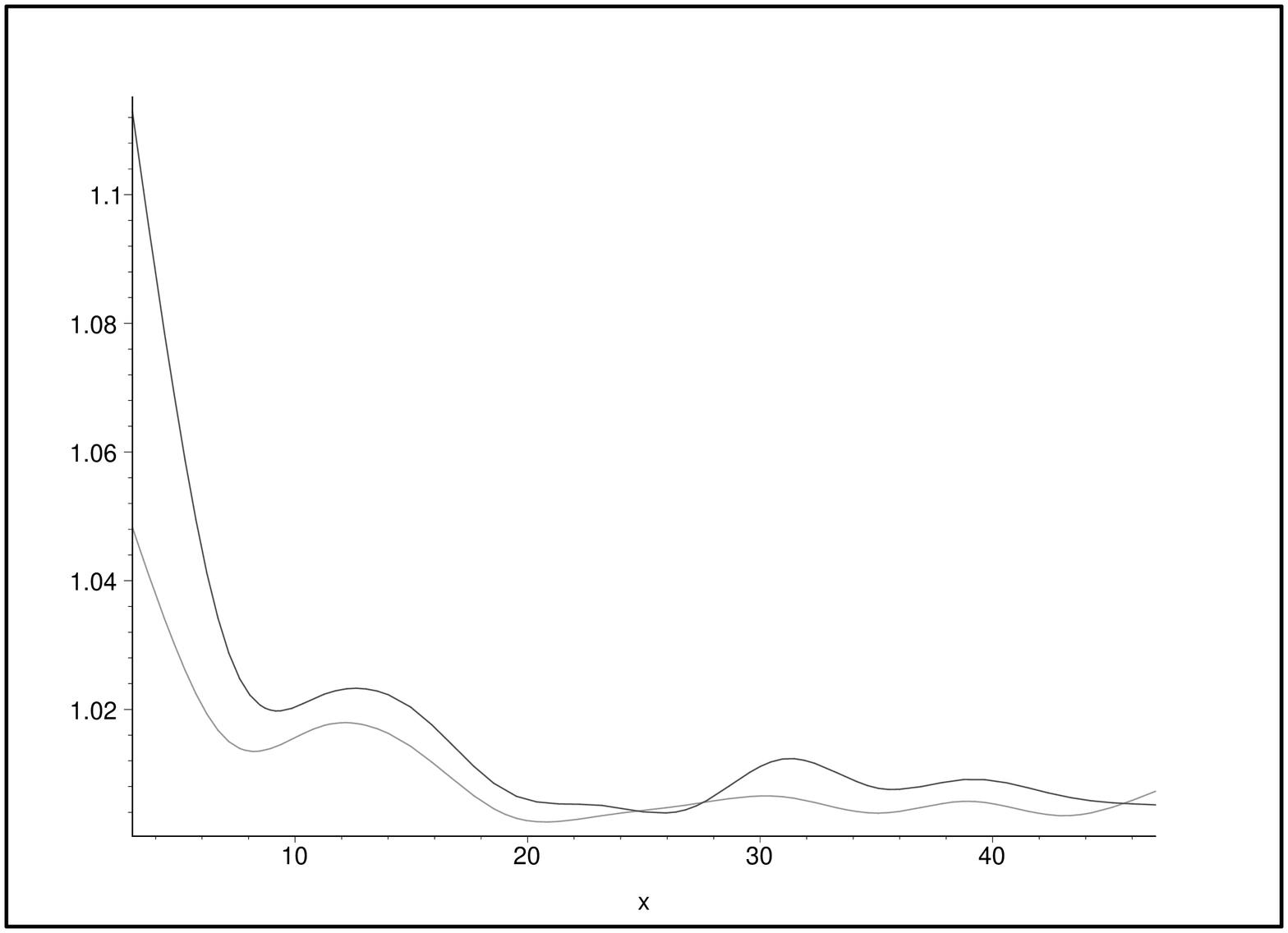}

\begin{caption}
{ Numerical solution and momodisperse approximation.
The number of droplets in the
units of the number predicted by the
mean values theory. }
\end{caption}
\end{figure}

Here there are two curves.
The curve with greater flash at
small $x=V$
corresponds\footnote{The volume and
the number of droplets in MVT have
a simple connection $1.28V =  N_{tot}$.}
to the precise solution, the
second curve comes from the monodisperse model.

We see that the monodisperse
model gives practically the
same behavior of the mean value of the droplets number
except the small $V$.

Numerical errors
can be evidently seen when we compare our
results with a
model where the first droplet appears
stochastically and all other droplets later appear
according MVT. The results are shown in
Figure 3 where
the lower curve comes from this model and
two upper curves have been
already drawn in the Figure 2.


\begin{figure}[hgh]

\includegraphics[angle=270,totalheight=10cm]{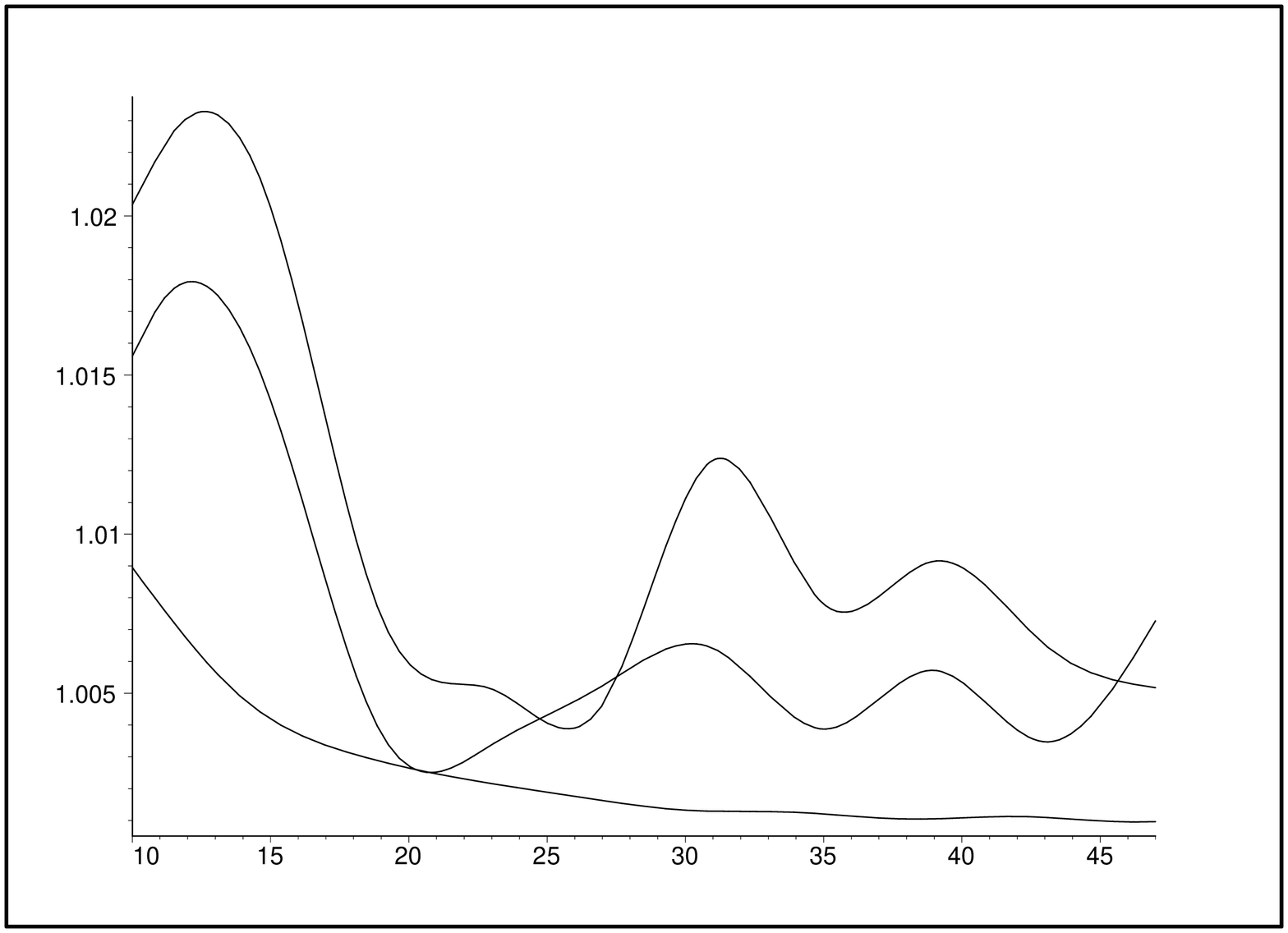}

\begin{caption}
{
Numerical solution.
The number of droplets in the units of
the number predicted by the mean values theory.
}
\end{caption}
\end{figure}

Certainly, the last model can
not give anything more that
 MVT (because the system will simply wait until the
 appearance of the
 first droplet and we can put $t=0$ at
 this very moment).
 Certainly there appears a little peak
 corresponding to the first droplet (That's why the
 dependence over
 volume $V$ of the system still exists.)
We see that here it is impossible to
analyze the
tails of dependencies (at big volumes),
they lies in the frames of error
of numerical
accuracy.

The effects of
discrete number of droplets can be seen from
the
following simple model: The first droplet appears
stochastically and later all other droplets
appear regularly
when the
integral of the nucleation rate over time attains
integer numbers.
Then results are shown in figure 4.


\begin{figure}[hgh]

\includegraphics[angle=270,totalheight=10cm]{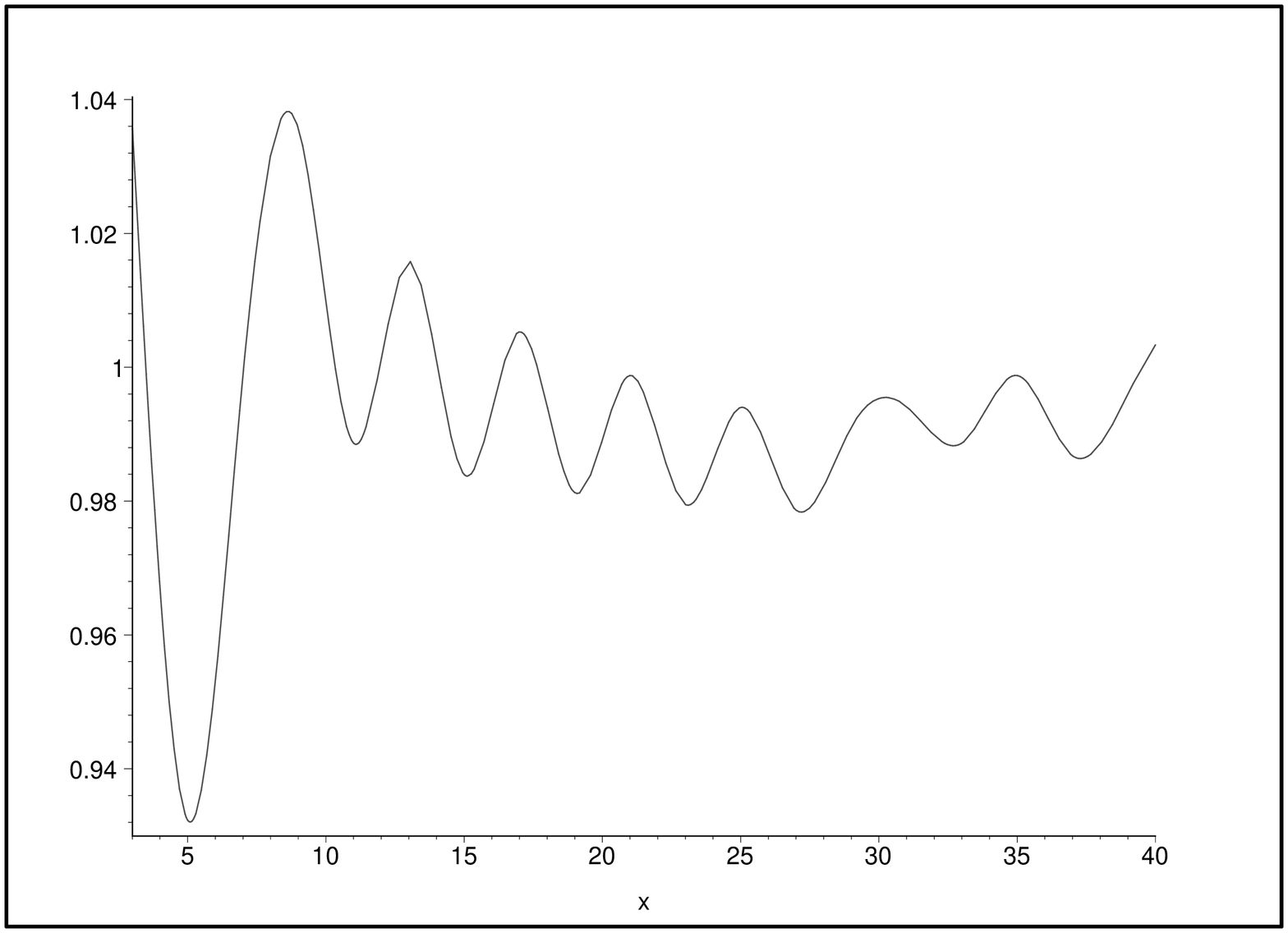}

\begin{caption}
{
Solution of the discrete regular model.
The number of droplets in the units of
the number predicted by the mean values theory.
}
\end{caption}
\end{figure}

We see that the deviations due to the
discrete character of
the
droplets number are even more essential than other
effects.
But in
reality
they do not take place because all $N_{eff}$
droplets appeared
independently and stochastically and these
shifts will compensate each another.

Now we shall turn to the
investigation of dispersion\footnote{We call the
relative square of dispersion  simply as dispersion.}.

The results
of dispersions for precise solution and for
monodisperse approximation  (for every model
there are two dispersions: one in
units
of $<N_{tot}>$ and another is in the units of
$<N_{tot}(V=\infty)>$,
but for such big values of $<N_{tot}>$
they practically coincide)
are shown in figure 5.


\begin{figure}[hgh]

\includegraphics[angle=270,totalheight=10cm]{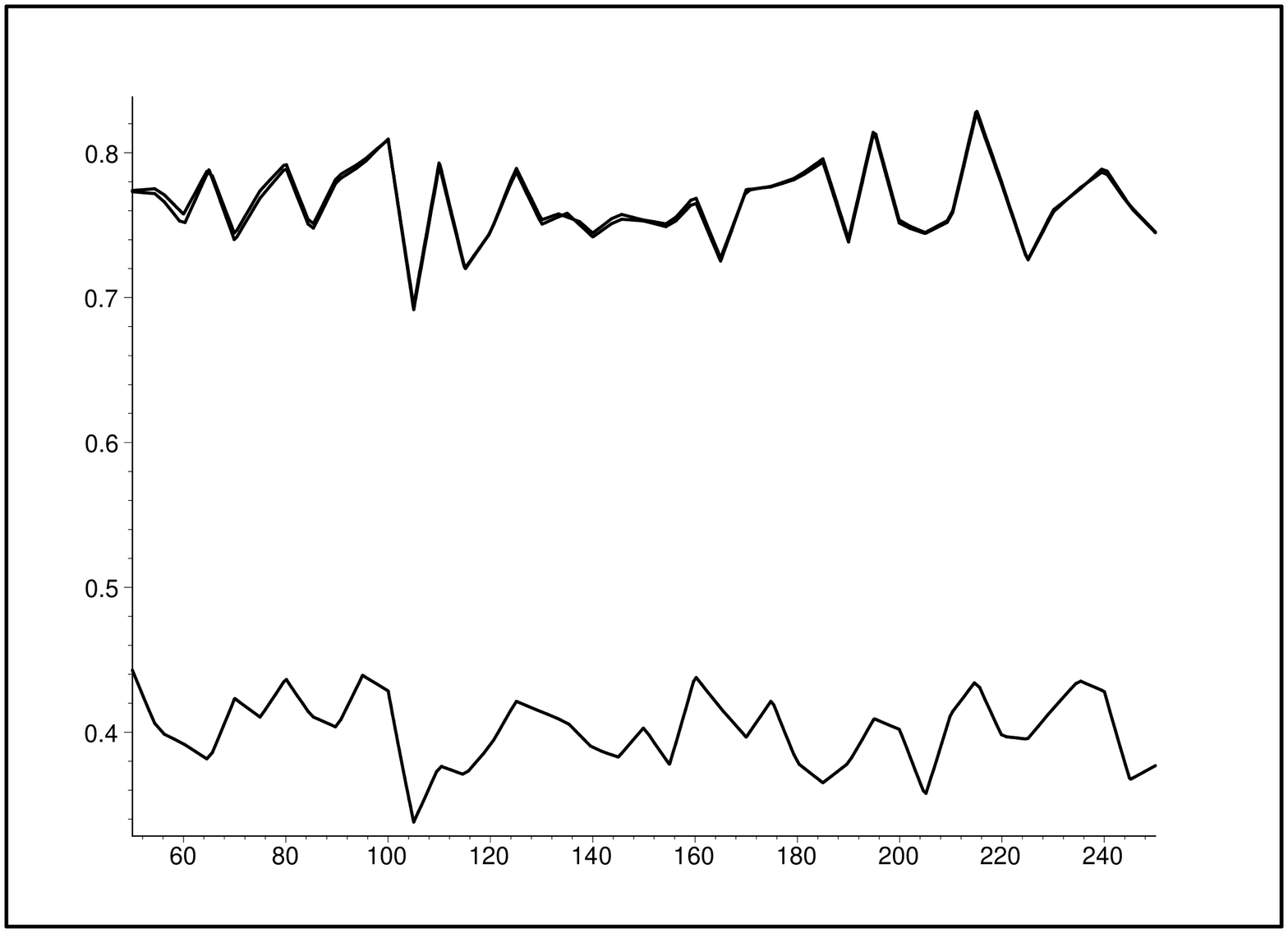}

\begin{caption}
{
Dispersions for precise solution (the upper curves)
and for the
monodisperse approximation (the lower curve).
}
\end{caption}
\end{figure}

The dispersion in the
monodisperse model can be easily
calculated. Really, the first part of droplets, i.e.
$$
N_{eff} =1/4
$$
has no dispersion
$$
D_{eff} = 0
$$
because the system simply waits until
there appears the $N_{eff}$ droplets precisely.
This behavior is
prescribed by the absence of the "external"
time in
the situation of decay. The system has only the
"internal time".
It waits until there will be enough
droplets to ensure the
beginning of the vapor consumption.
This approximate
picture is more realistic than a picture
with  fixed moment of the
boundary between the cycle of
appearance of the main consumers
and the cycle of the
real intensive consumption of vapor.

The droplets appeared
during the cycle of consumption are
born under the
known supersaturation and, thus, under the
known rate
of nucleation. So, we have the free appearance
of droplets with
known number of possible acts of appearance
or the appearance with a known time lag
of the cut-off. So,
the dispersion
will be equal to the dispersion of
appearance of the free droplets with a mean value
$$
<N_{rest}> =
<N_{tot} - N_{eff}> = 1.28 - 0.25 = 1.03
$$
So, the $D$ is
$$
D_{rest} = 2 <N_{rest}>
$$
The total
dispersion (after the  combination of gaussians)
will be found from
$$
D_{total} = D_{eff} + D_{rest} = 1.03
$$
Being referred to the standard $D$ which is two
total numbers of droplets
$D_{standard} = 2 N_{tot} = 2*1.28$
it gives the relative dispersion
according to the following
expression
$$
D_{rel} = 1.03/1.28 = 0.75
$$

We see that the result
for dispersion of the monodisperse
approximation
isn't too close to the real solution. What is
the matter of this
discrepancy? Really, we see that the
monodisperse
approximation which has been used (let us call
it the "standard
monodisperse approximation") has some
disadvantages:
\begin{itemize}
\item
Already all essential
droplets appear at the initial moment
of time. It leads to the
inequality of the real mean value
of "essential" droplets and
the coordinate of monodisperse
peak.
\item
We have to keep the balance of substance but
since the mean coordinate (size) of
effective droplets is less
than the coordinate of
monodisperse peak then the number of
essential droplets in
monodisperse approximation have to be
less than the real
number of essential droplets. This shows
why the real dispersion is less
than the result given by the
standard monodisperse model.
\end{itemize}

So, we have to put
the coordinate of monodisperse peak to
the real mean coordinate of "essential" droplets.
Then
we come to the following
model

\begin{itemize}
\item
At $z=l$ the
monodisperse peak is formed. It contains
$N_{eff} = 1*l + 1*l$ droplets
(here it is taken into
account that the amplitude here
is unperturbed and it
equals to $1$ and
the peak has rather symmetric form).
\item
The total number of droplets is calculated as
$$
N_{tot} \approx  N_{tot\ 1} =
l+ \int_0^{\infty} \exp(-2 l x^3) dx
$$
or
$$
N_{tot} \approx  N_{tot\ 2} =
2l+ \int_l^{\infty} \exp(-2 l x^3) dx
$$
\end{itemize}

This model will be called as
the "primitive monodisperse
model".

Unfortunately this model
can not give the correct value of
the total number of droplets.
The value $N_{tot}$ as a
function of $l$ is drawn in figure 6.


\begin{figure}[hgh]

\includegraphics[angle=270,totalheight=10cm]{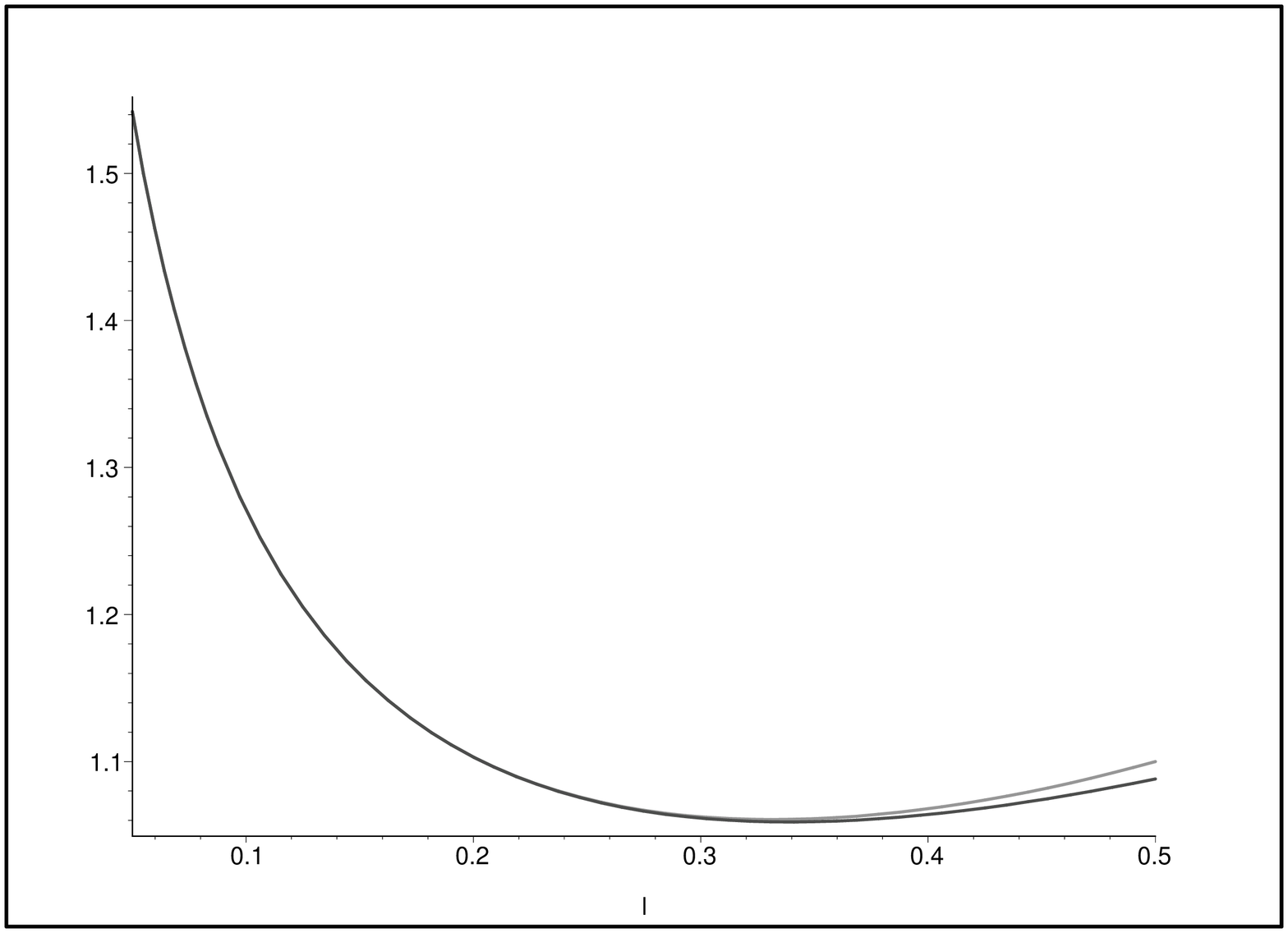}

\begin{caption}
{
The total number of
droplets as a function of parameter $l$
in a primitive monodisperse model.
}
\end{caption}
\end{figure}

Here two curves
for $N_{tot\ 1}$ and for $N_{tot\ 2}$ are
drawn. We take the relative values
referred to the precise value $N_{tot} =
1.28$.
They are so close that one can not separate them.
Only at the
tail one can see the thick line corresponding
to the little
deviation $N_{tot\ 1}$ from $N_{tot\ 2}$.

We see that
the result is greater than $1.1$ even in the minimum
corresponding to $l= 0.33$.
Namely this value is the most
suitable value of parameter $l$ in
the primitive monodisperse
model.

The value of  minimum of the
droplets number is important
not only because it is the
closest number to the real value
$1.28$ but also because it is the
minimum and the minimum
in the droplets number corresponds
to the minimum of the
free energy of the total system.
So, this property can be
effectively used and we shall
seek in future the values of
such  minima in more sophisticated models.

The evident weak feature of this
model is that the position
of the monodisperse peak is put
directly in the middle of
the period of formation of all effective
droplets. It
supposes the relative symmetry
of the vapor consumption of
all effective droplets.
Certainly it isn't true, one can
see on the base of the first iteration that the
subintegral function resembles $(z-x)^3$ and isn't
symmetric.
So, now we shall introduce the arbitrary shift
of the position
of the peak and formulate the next model as
following:

\begin{itemize}
\item
The lenght
of monodisperse peak is $z=2*l$. It contains
$N_{eff}
= 1*l + 1*l$ droplets (here it is taken into
account that
the amplitude here is unperturbed and it
equals to $1$).
\item
The position of the peak is formed at $z=2*l - b$
with the parameter $b$. Earlier it was $b=l$.
\item
The total number of droplets is calculated as
$$
N_{tot} \approx  N_{tot\ 1} =
2l-b+ \int_0^{\infty} \exp(-2 l x^3) dx
$$
or
$$
N_{tot} \approx  N_{tot\ 2} =
2l+ \int_{b}^{\infty} \exp(-2 l x^3) dx
$$
\item
The value of $b$ is
determined to touch the value $1.28$ at
the minima over $l$

\end{itemize}

This model will be called as the
"advanced monodisperse model".

The calculations
give $b=0.33$ and the function $N_{tot}$
over $l$ at $b=0.33$ is drawn in figure 7


\begin{figure}[hgh]

\includegraphics[angle=270,totalheight=10cm]{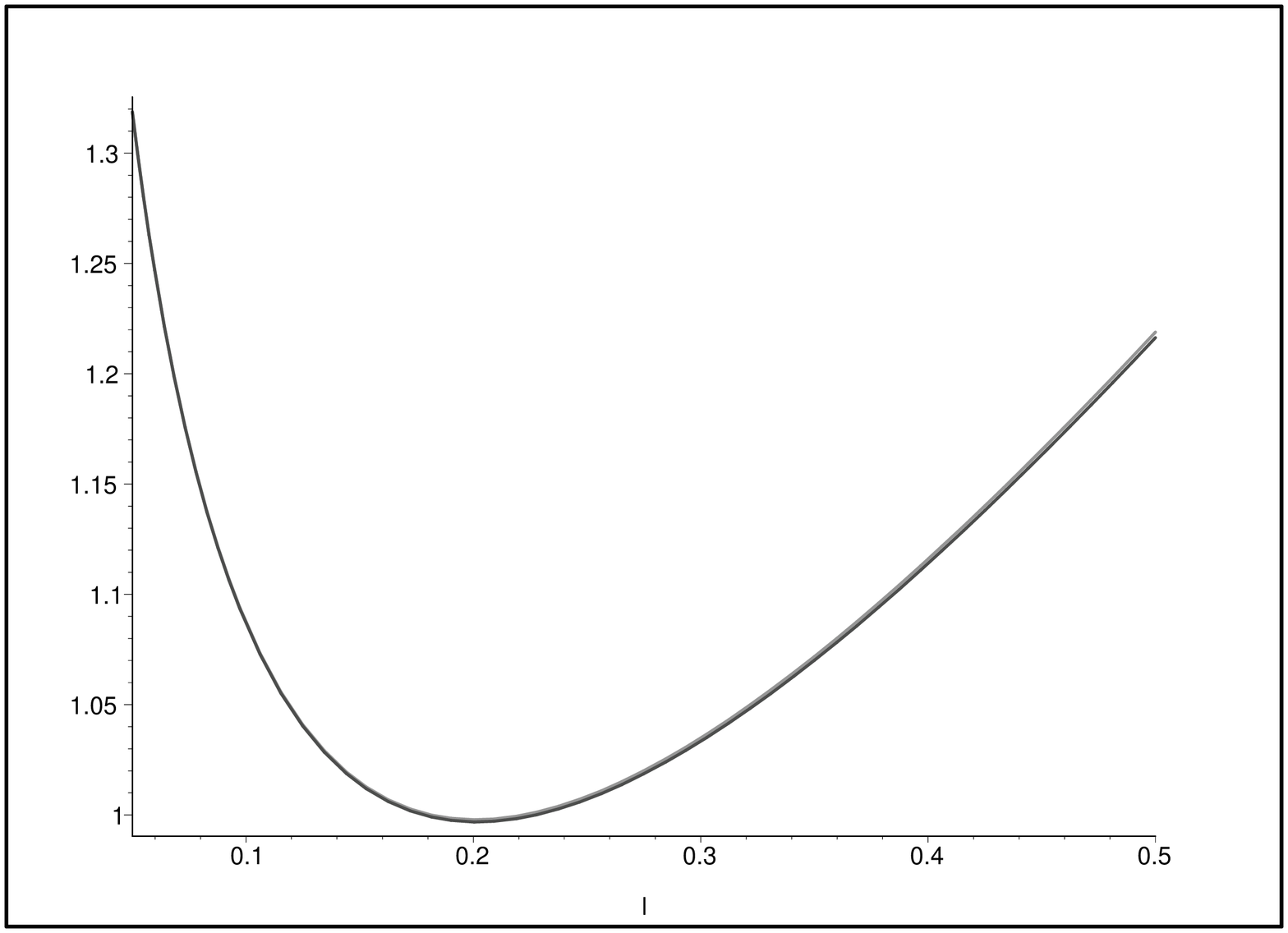}

\begin{caption}
{
The total number of
droplets as a function of parameter $l$
in the
shifted monodisperse model at $b=b_0 \equiv 0.336$.
}
\end{caption}
\end{figure}

Here also two curves with the same
meaning coincide.

The number of effective droplets here is $N_{eff} = 0.4$.

Theoretical result
originally going from the floating
monodisperse approximation (see \cite{Monodec})
is
$$
b = 2l - (1/4)*2l = 0.3
$$
Here $1/4$ is the
same as $N_{ess}$ in our first model.

Now we shall clarify the physical sense
of this
approximation. One can see that
$$
\int_0^b \rho^3 d\rho
\approx
 \int_b^{0.4} \rho^3 d\rho
$$

We  can require
that the quantity of substance in the "left"
part of monodisperse
spectrum (i.e. in the droplets
appeared before the peak)
equals to the quantity of
substance in the
"right" part of spectrum (i.e. in the
droplets appeared
after the peak of spectrum).
Then we  come to the
practically same results as before.

It is very easy to get the
results for dispersion in this
model.
$$
<N_{rest}> =
<N_{tot} - N_{eff}> = 1.28 - 2*l  = 0.88
$$
So, the dispersion is found from
$$
D_{rest} = 2 <N_{rest}>
$$
The total dispersion
(after the combination of gaussians) is
given by the following expression
$$
D_{total} = D_{eff} + D_{rest} = 0.88
$$

Being referred to the
standard dispersion which is two
total numbers of droplets
it gives the relative dispersion according to
$$
\gamma = 0.88/1.28 = 0.68
$$
This value is rather close to the result of computer
experiments. Then we can fulfill some
formal summations
analogous
to \cite{statiae} due to the self similarity of the
process
of
nucleation (see \cite{varios}) and get the result
which
practically coincides with numerical simulation.

Now we shall turn
to numerical simulation for the last  model.

The mean number of
droplets is drawn in Figure 8.


\begin{figure}[hgh]

\includegraphics[angle=270,totalheight=10cm]{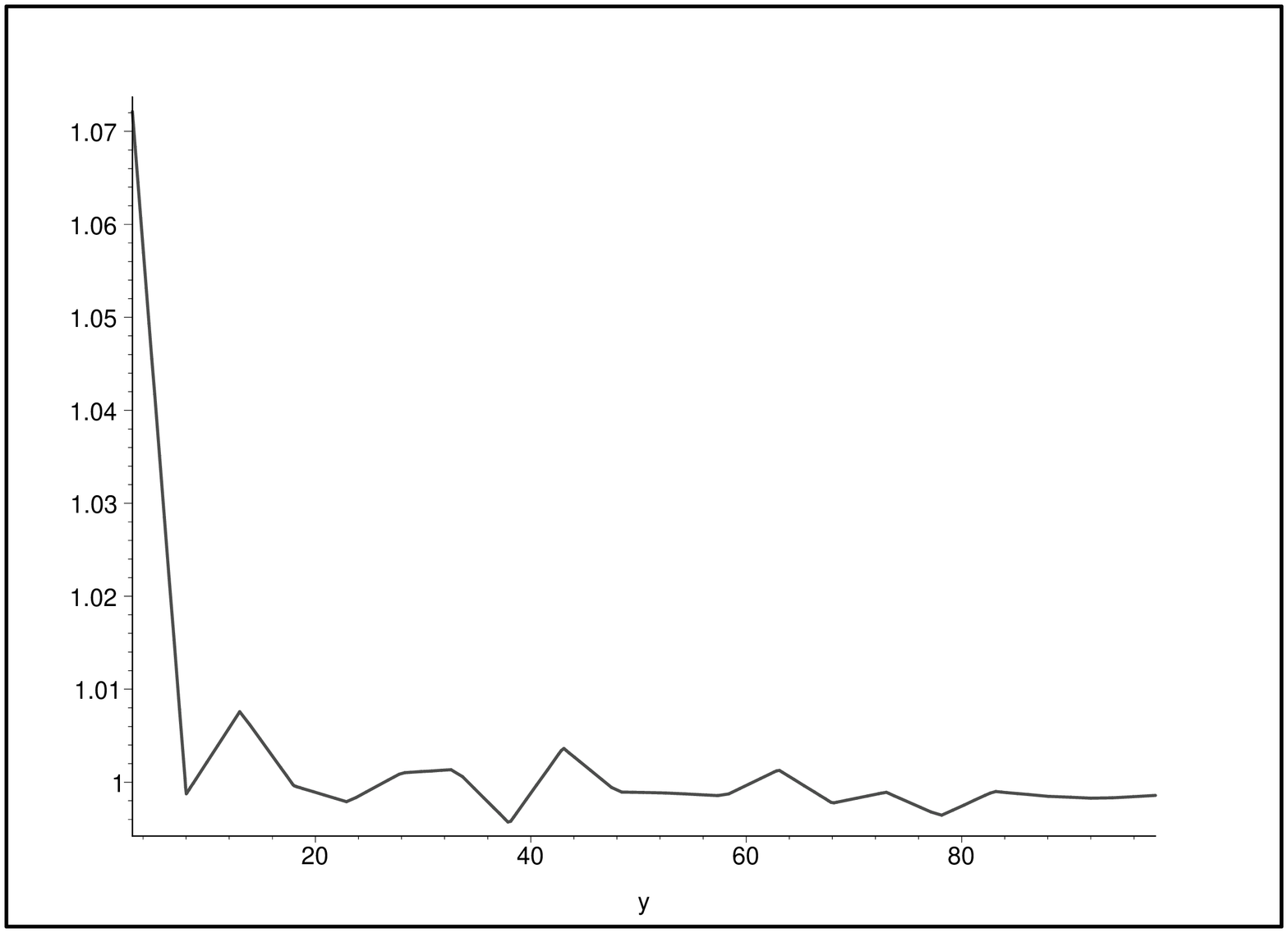}

\begin{caption}
{
The mean number of
droplets as a function of the volume of
the system.
}
\end{caption}
\end{figure}

The dispersions are drawn in Figure 9.

\begin{figure}[hgh]

\includegraphics[angle=270,totalheight=10cm]{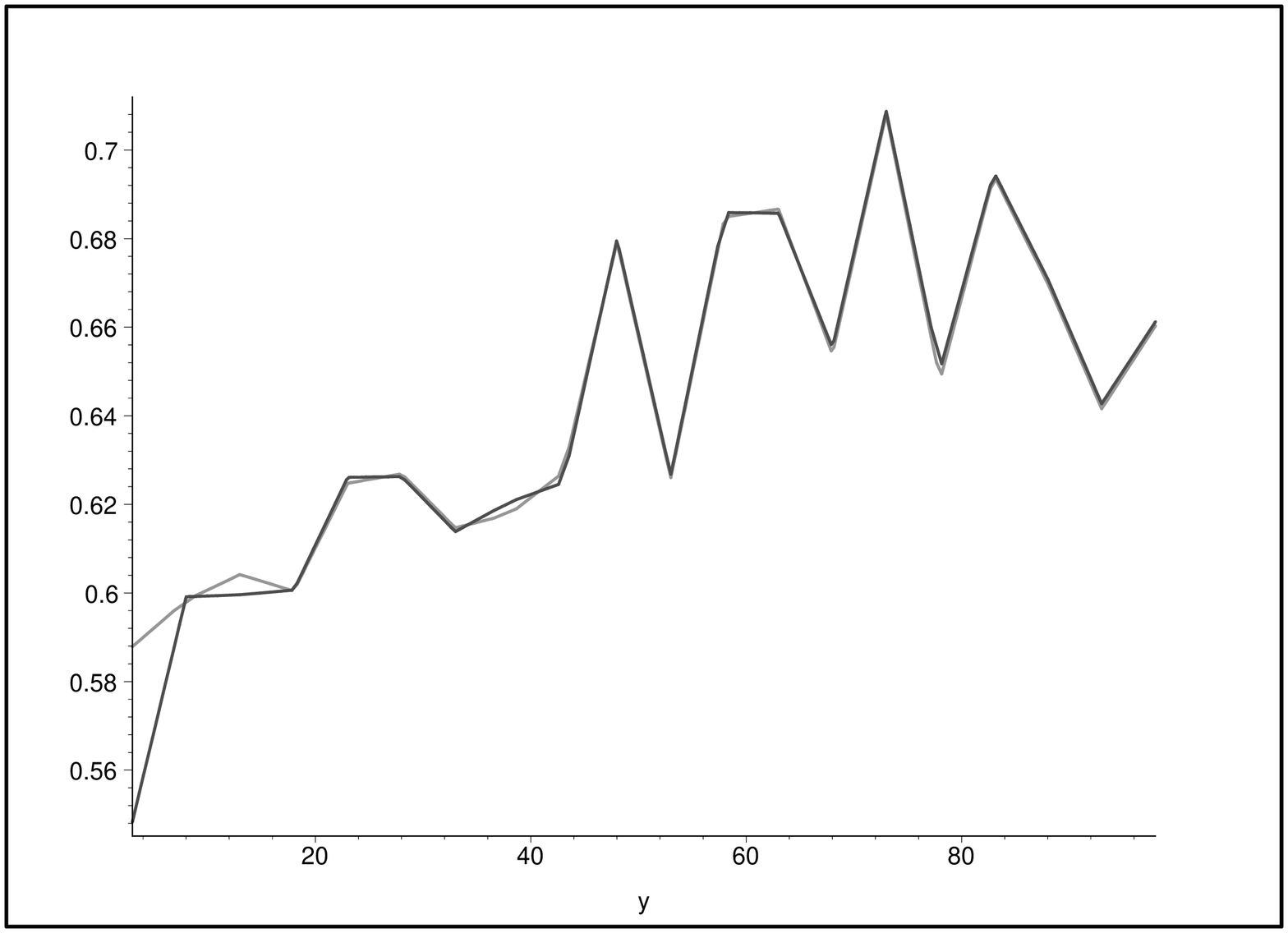}

\begin{caption}
{
Relative dispersions as  functions of the volume of
the system.
}
\end{caption}
\end{figure}

We see that  numerical results of this
model are practically the same as the
results of computer simulation of the
process in initial formulation.

The use of
monodisperse approximation in investigation of
stochastic
effects of nucleation has certain advantages.
The first is the real
simplicity of this method. Really,
 two summations and
 multiplications  lead to a final result.
But this
simplicity isn't
only a reduction of amount of
calculations.
Behind this simplicity lies the real physics
of manifestation of
stochastic effects in decay of
metastable state.

The process
of decay can be qualitatively described as
following.
The system waits for
formation of necessary amount of
droplets which later
will be the main consumers of vapor in
the whole nucleation
period.
Later all these droplets
will be the main consumers of all
surplus metastable phase.
Every droplet consumes
approximately equal amount of
metastable phase.

The process of phase
transition has a three cycle
structure.
In the general  period of
the whole phase transition
the period of formation of
the main consumers of vapor,
i.e. the nucleation period,
can be extracted. In the
nucleation period the
sub-period of formation of the main
consumers of vapor at
nucleation period, i.e. the initial
period of nucleation, can
be extracted. The further
extraction doesn't
take place. As it has been shown
the system simply waits
until formation of the necessary
number of droplets
characteristic for the initial cycle.
The initial cycle can not be
further divided, at least in
frames of description of
stochastic effects by means of
dispersion and the mean
value of the droplets number.

Certainly, this picture is
approximate and in  some rare
cases it can be violated.
In principle it is possible that
the total number
of droplets  is less even than $N_{ess}$.
But since
$N_{ess}$ is seriously less than $N_{tot}$ the
probability
of such event is extremely low. Account of
dispersion of
initial period is analogous to \cite{explanation}
where it was
done for the smooth behaviour of external
conditions.
Also one can found it in \cite{statiae} but
since there
the model with a fixed boundary is studied,
some evident
modifications have to be performed.

When we are
interested in more specific characteristics
such as
higher momenta of the
differential distribution function
then one has to
fulfill
quite the same extraction
of sub-sub-periods in initial
sub-period. Practically nothing will
be changed. So, one can
observe the chain of sub-periods responsible for
deviations in values of high momenta of distribution.
This chain is
limited by the formation
of the first droplet.

In some
regimes of the droplets
growth one can observe the direct
influence of the moment
of formation of the first droplet on
the nucleation
kinetics. This picture will be published
separately.

Now we shall  turn to investigation of
the diffusion regime of droplets
growth. This consideration is rather
formal one because  kinetics of
nucleation in such a regime is based on
some other features (see
\cite{PhysicaA}). But we  still perform
calculations to see  specific features
of stochastic effects in this case.

To investigate the process
of decay in diffusion regime of
droplets growth one can also use
the modified monodisperse
approximation.

To get the monodisperse
approximation  we consider equation
$$
g(z) = \int_0^z (z-x)^{3/2} \exp(-g(x)) dx
$$
The remormalization with the absence of coefficient
corresponds to more
natural expression for the number of
droplets
in monodisperse peak.

The number of
droplets in first iteration is given by
$$
N_1 = \int_0^{\infty}
\exp(-\frac{2}{5} x^{5/2} )
dx =1.27
$$

Traditional monodisperse
approximation requires to put the
monodisperse
spectrum at $z=0$. The number of droplets
$N_{mono}$
in
the monodisperse peak is chosen to satisfy the first
iteration and looks like
$$
N_{mono} =
(2/5) z
$$
It corresponds to the
value of the boundary $p=0.4*1.25$ between the
cycle of formation of
the main consumers during the period
of nucleation and the
rest droplets. Here $1.25$ is the
charateristic lenght
of spectrum in this renormalization.

In this approximation
the dispersion will be calculated
according to
$$
D= 2 (1.27 - 0.4)/1.27 * N
$$
where $N$ is the
mean value of droplets. So, the relative
dispersion will be
$$
\gamma = (1.27 - 0.4)/ 1.27 = 0.685
$$

This value is too far from the real value
but still
this value is closer to the real value
$0.45$ of dispersion than
the result given by the direct
application of
recipe given in \cite{Koll} to the diffusion
regime.

Now we shall use more
realistic monodisperse approximation.
Why the coordinate of monodisperse
spectrum is put to
$z=0$? Certainly, there is no
strong motivation of such
choice except
notation that $z=0$ corresponds to the
maximum value of
spectrum and the maximum value of subintegral
function in expression for $g$.

As one of contre-arguments
one can say that $z=0$ leads to
the absolutely
unsymmetric monodisperse spectrum. So, it
isn't too reasonable to choose $z=0$.

Instead of this choice
we shall leave the coordinate of
spectrum as a free parameter.
Namely we suppose that in
the monodisperse
spectrum there is $2 * l$ droplets.
since $l$ is small
in comparison with $1.25$ one can state
that these droplets
were formed at ideal
supersaturation.
Then the upper boundary of the region of
formation of monodisperse
peak is $2l$. We suppose that the
monodisperse spectrum is
formed at $2l - b$. Here $b$ is a
free parameter.

The total number of droplets is calculated as
$$
N_{tot} = 2l + \int_b^{\infty} \exp(-2lz^{3/2})
dz
$$

In addition we have to
suggest a recipe to determine  parameter $b$.
One can prove that at arbitrary $0<b<2l$ the value
$N_{tot}$ as a function of $l$
has minimum. It is clear that
this minimum is the most
profitable from the energetical point
of view. So, the
real evolution corresponds to the choice of
$l,b$  giving minimum of $N_{tot}$.

We shall require
that the value of munimum has to be equal
to the
real number of droplets. One can use as this number
the number of droplets
in the first iteration.

Calculations give
$l=0.35$, $b=0.65$.
So, dispersion will be
$$
\gamma = (1.27 - 2*0.35)/1.27  =0.448
$$
This value coincides with the result of numerical
simulation.

One can note that the value  $N_{tot}$ as $1.27$
which is given by the
first iteration corresponds to the similarity of
spectrum. Namely this
similarity was already used to refine
the results of the two cycle model.

We see that the value $2l$ of the
"length" of essential part of spectrum
is more than $60$ percents  of the
lenght of th etotal spectrum. It means
that the majority of droplets
participates  in vapor consumption
during the nucleation period. This
corresponds to another structure of
nucleation kinetics presented in
\cite{PhysicaA}.



\end{document}